\documentclass[
  prb,
  twocolumn,
  final,
  floats,
  floatfix,
  amsmath,
  amssymb,
  nofootinbib,
  byrevtex,
]{revtex4}

\usepackage{graphicx}

\newcommand{\rr}[0]{\mathbf{r}}
\newcommand{\pp}[0]{\mathbf{p}}

\newcommand{\ud}[0]{\mathrm{d}}
\newcommand{\abbr}[1]{{\texttt{#1}}}

\begin{document}
\title{%
Conformational free energies of methyl-$\alpha$-L-iduronic
and methyl-$\beta$-D-glucuronic acids in water.
}
\author{Volodymyr \surname{Babin}}
\author{Celeste \surname{Sagui}}
\affiliation{
Center for High Performance Simulations (CHiPS) and
Department of Physics, North Carolina State University, Raleigh,
NC 27695
}
%
%
\begin{abstract}
We present a simulation protocol that allows for efficient sampling of
the degrees of freedom of a solute in explicit solvent. The protocol
involves using a non-equilibrium umbrella sampling method, in this case
the recently developed adaptively biased molecular dynamics method, to
compute an approximate free energy for the slow modes of the solute in
explicit solvent. This approximate free energy is then used to set up a
Hamiltonian replica exchange scheme that samples both from biased and
unbiased distributions. The final accurate free energy is recovered via
the \abbr{WHAM} technique applied to all the replicas, and equilibrium
properties of the solute are computed from the unbiased trajectory. We
illustrate the approach by applying it to the study of the
puckering landscapes of the methyl glycosides of $\alpha$-L-iduronic
acid and its C5 epimer $\beta$-D-glucuronic acid in water. Big savings
in computational resources are gained in comparison to the standard
parallel tempering method.
\end{abstract}
\maketitle
%
\section{\label{sec:intro}
Introduction
}
%
The field of \textit{glycobiology} has undergone rapid growth since its
name was coined in 1988\cite{Rademacher_T_88} to refer to the study of
the molecular structure and biology of saccharides (or glycans) as they
happen in the cell. Linear polysaccharides are ubiquitous in nature.
Glycoproteins, proteoglycans and glycolipids are the most common
glycoconjugates in mammalian cells (found mainly in the outer cell wall
and fluids). In particular, glycosaminoglycans (heparin, heparan
sulfate, chondroitin, keratin, dermatan, and many others) comprise a
disaccharide
unit formed by two aminosugars linked in an alternating fashion with
uronic acids\cite{Essentials_of_Glycobiology}. In this work we consider
L-iduronic acid (IdoA), which is the major uronic acid component of the
glycosaminoglycans dermatan sulfate and
heparin\cite{Essentials_of_Glycobiology}. We also consider its C5
epimer, D-glucuronic acid (GlcA), that dominates in chondroitin
and hyaluronan.
The IdoA is an uncommon molecule among the pyranoses (carbohydrates whose
chemical structure is based upon a six-membered ring of five carbon atoms
and one oxygen) because it adopts several conformations, and therefore
it is believed to
increase the flexibility and conformational
freedom of the corresponding polysaccharide chains\cite{Ferro_1990}.

Computational studies of polysaccharides have evolved 
along several directions. In particular, considerable work has been
devoted to quantum chemical studies of the energetics of relatively
small compounds (see, for example, Ref.\onlinecite{book:Rao} and references
therein). While very accurate, such \textit{ab-initio} calculations typically
cannot handle explicit solvent over the nanosecond time scale. Yet, it is
known that solvation plays a crucial role in the conformational
preferences of carbohydrates\cite{Kirschner_K_01,Almond_A_03}.
Until recently, only coarse-grained models were able to breach the long
time scales involved in conformational sampling.  However, methodological
advances coupled to growing computational power have allowed the realization of
all-atom,
explicit solvent classical molecular dynamics over hundreds of
nanoseconds\cite{Krautler_V_07}.
This of course requires the development and improvement
of force-fields\cite{glycam06}, which need to be validated by comparison
with experiments. Reliable validation necessitates an accurate determination of 
the properties of the compounds in
explicit solvent\cite{Christen_M_08}, that can be computationally very
demanding.

Often, the dynamical aspect of the trajectory generated in the
course of a molecular dynamics (\abbr{MD}) simulation is not of interest
\textit{per se},
and the trajectory is used to study an ensemble
statistics of some quantities. It was realized long
time ago that \abbr{MD} is quite limited when employed for sampling
purposes because the trajectory gets trapped in the vicinity of the free
energy minima and is very unlikely to cross barriers higher than a few
$k_BT$. Many approaches to enhance the sampling have been proposed, with
different measures of success. It is beyond the scope of this paper to review
the different methods  (the reader can find a review in
Ref.\onlinecite{Christen_M_08}), but  we will briefly go over the two
methods employed in this work, namely the ``replica exchange''
method\cite{Geyer_C_91} and the adaptively biased molecular dynamics
method\cite{Babin_V_2008}.

The idea behind the replica exchange method is to consider several \abbr{MD}
trajectories that are generated using different Hamiltonians and
exchange those Hamiltonians (or trajectories) in a way that preserves
detailed balance so that all the Boltzmann distributions that correspond
to the Hamiltonians involved get sampled simultaneously in a mutually
enhancing way. The overall performance of the method is determined by
the ``right'' choice of the Hamiltonians involved in the construction.
For example, ``parallel tempering'' (the most popular incarnation of the
method) builds the family of Hamiltonians by re-scaling the original
Hamiltonian. This is identical to the situation where one Hamiltonian is
used at different temperatures -- high-temperature replicas have higher
barrier-crossing rates and continuously ``seed'' the low-temperature
replicas with different configurations. The advantage of this method is
that it is very general requiring little prior knowledge about the
system. The disadvantage is that, as the number of degrees of freedom
increases, more and more replicas are required to cover the desired
range of temperatures while maintaining satisfactory exchange rates. In
the pure parallel tempering case, the number of replicas has been shown
to grow as the square root of the number of degrees of
freedom\cite{Kofke_D_2002,Kofke_D_2004}.

Over the past few years, several methods targeting the
computation of the free energy using non-equilibrium simulations
have become popular.
These methods all estimate the free energy as a function
of ``collective variables'' from an ``evolving'' ensemble of
realizations\cite{Lelievre_T_2007,Bussi_G_06}, and use that estimate to bias
the system dynamics, so as to flatten the effective free energy surface.
One of the first methods that used the idea of a dynamically evolving
potential that forces the system to explore unexplored regions of the
phase space was the local elevation method\cite{Huber_T_94}.
 More recent developments include the
adaptive force bias method\cite{Darve_E_01}, the Wang-Landau
approach\cite{Wang_F_01}, and the
non-equilibrium metadynamics\cite{Laio_A_02,Iannuzzi_M_03} method.
Collectively, all these methods may be considered as umbrella sampling
methods in which the biasing force compensates for the free energy gradient.
In the long time limit, the biasing potential eventually reproduces the
previously unknown free energy surface.
The adaptively biased molecular dynamics (\abbr{ABMD}) 
method\cite{Babin_V_2008} grew out
of our efforts to streamline the metadynamics method
 for classical
biomolecular simulations\cite{Babin_V_2006}. It shares all the important
characteristics of the above methods while including enhancements
in its own right such as a favorable $O(t)$ scaling
with time $t$, and a fewer number of control parameters.
Adaptive bias methods have been recently used to enhance the
conformational sampling of small sugars in explicit water. These efforts
include the use of the local elevation method to calculate the relative
free energies and interconversion barriers of the ring conformers
of $\beta$-D-glucopyranose in water\cite{Hansen_H_2009}, and the use of
metadynamics to recover the conformational free energy surface of
$\alpha$-N-Acetylneuraminic acid\cite{Spiwok_V_09} and
$\beta$-D-Glucopyranose\cite{Spiwok_V_10}. The latter work
is an important contribution to the force-field validation
process.

It is possible to harness the power of replica exchange with
a suitable family of Hamiltonians that are built by biasing
the original Hamiltonian
along relevant slow modes. These biasing potentials
can be readily computed using the adaptive bias methods.
In this paper, we combine the \abbr{ABMD} method
with the replica exchange and the weighted histogram analysis
(\abbr{WHAM}\cite{Ferrenberg_A_89,Kumar_S_92}) methods
in order to achieve robust sampling of the conformational space
of the pyranose ring in explicit solvent, and recover the
potential of mean force (free energy) associated with two
intra-ring dihedral angles describing the slow puckering modes.
The free energy alone could be computed
by ``correcting" the biasing potential obtained by \abbr{ABMD}
with an equilibrium run, as discussed at length in
Ref.\onlinecite{Babin_V_2006}. The approach
explored in this paper replaces the ``correction" step with
a replica-exchange simulation that both ``corrects" the approximate
biasing potential \textit{and} efficiently samples unbiased
equilibrium configurations. 

Our study has two main goals. First, we present a simulation
protocol that allows for efficient sampling of the degrees of freedom
of a solute in explicit solvent. Second, we illustrate how this approach
can be used to validate the new \abbr{GLYCAM~06}
force field\cite{glycam06} by means of applying it to the study of the
puckering landscapes of the methyl glycosides of $\alpha$-L-iduronic acid
and its C5 epimer $\beta$-D-glucuronic acid (see Fig.\ref{fig:epimers}).

The paper is organized as follows: the methodology is described in
the next section which is followed by the presentation of the
simulation protocol along with the results. The paper then ends with a
short summary.
%
\section{\label{sec:methods}
Methods
}
%
Let us consider a system of $N$ classical particles. The state of the system
can be described by the positions of the particles $\rr_1,\dots,\rr_N$
and their momenta $\pp_1,\dots,\pp_N$. Let
\begin{equation}\label{eq:Hamiltonian}
  \mathcal{H} = \sum\limits_{a=1}^{N}\frac{\pp^2_a}{2m_a}
      + \Phi(\rr_1,\dots,\rr_N)
\end{equation}
be the Hamiltonian of the system. Here, $m_a$ are the masses of the
particles and $\Phi(\rr_1,\dots,\rr_N)$ is the potential energy. In the
following, we omit the atomic indices if this does not lead to
confusion, that is, $\rr$ represents $\rr_1,\dots,\rr_N$ and $\pp$
represents $\pp_1,\dots,\pp_N$.

A routine task in  molecular modeling involves sampling
configurations according to the canonical distribution
\begin{equation}\label{eq:canonical}
  \mathcal{P}(\pp,\rr) \propto e^{\textstyle -\mathcal{H}(\pp, \rr)/k_BT}
\end{equation}
($k_B$ is the Boltzmann constant and $T$ is the temperature). One way of
carrying out such sampling is, for instance, to generate dynamics
via the Langevin equation such that its limiting probability
distribution is the desired equilibrium distribution\cite{Brunger_A_84}.
Unfortunately this basic strategy often fails because the system
gets either trapped in the neighborhood of a potential energy
minimum or locked in some region of  phase space due to
entropic bottlenecks. Hence, other more
elaborate approaches have been proposed throughout the years
(see, for example, a review in Ref.\onlinecite{Christen_M_08}).
In this paper we describe how a combination of the replica
exchange\cite{Geyer_C_91}, \abbr{ABMD}\cite{Babin_V_2008} and
\abbr{WHAM}\cite{Ferrenberg_A_89,Kumar_S_92} methods
can be used to enhance sampling, and illustrate the use of this protocol by
applying it to the
study of small solutes in explicit solvent. For clarity and
completion, we briefly review these methods below.
\subsection{Replica exchange method}
The replica exchange idea\cite{Geyer_C_91} consists of considering
several Hamiltonians,
$\mathcal{H}_{a}(\pp,\rr)$, $a = 1,\dots,M$ (with the original
$\mathcal{H}(\pp,\rr)$ being of one them), running the dynamics for each of
them, and trying to ``cross-pollinate'' these dynamics by exchanging the
trajectories between the different Hamiltonians from time to time. More
formally, the method can be implemented as follows: once in a while a
random pair of replicas is chosen and an exchange is attempted with the
acceptance probability carefully tuned to maintain  detailed balance
in the extended ensemble\cite{Geyer_C_91,Sugita_Y_2000}
\begin{equation}\label{eq:exchange_probability}
  \mathcal{P}_{ab} =
  \left\{%
    \begin{array}{ll}
    1, & \Delta\leq 0,\\
    e^{\textstyle -\Delta}, & \Delta > 0,
    \end{array}
  \right.
\end{equation}
\begin{eqnarray*}
  \Delta &=& \frac{1}{k_BT_a}\left[\Phi_a(\rr^b) - \Phi_a(\rr^a)\right]\\
         &+& \frac{1}{k_BT_b}\left[\Phi_b(\rr^a) - \Phi_b(\rr^b)\right],
\end{eqnarray*}
where $T_{a,b}$ denote the temperatures of the replicas and only the
potential energy is shown (the momenta are integrated out). In practice,
the momenta get simply rescaled upon a successful exchange. In this
scheme, high, inaccessible barriers in one replica can be
smaller in another replica and hence this random walk over the replicas
facilitates barrier crossings.
Care must be taken with the choice of Hamiltonians since it determines
the performance of the method. There are many
variations of replica exchange, with the most popular one being
``parallel tempering'': the family of the Hamiltonians is constructed
from the rescaled copies of the original one (this is equivalent to
having the same Hamiltonian with different temperatures). Parallel
tempering has the advantage of being quite general, but the applicability
of the method can be limited by the fact that the number of replicas
needed to cover a desired temperature range (while maintaining an
efficient random walk across those temperatures) grows as the square
root of the number of degrees of
freedom\cite{Kofke_D_2002,Kofke_D_2004}.
\subsection{Adaptively biased molecular dynamics}
The \abbr{ABMD}\cite{Babin_V_2008} method is an umbrella sampling method with
a time-dependent biasing potential that can be used to compute the Landau
free energy of a collective variable $\sigma(\rr)$,
which is understood to be a smooth function of the atomic positions.
The Landau free energy is defined as\cite{Frenkel}:
\begin{equation}
  \nonumber
  f(\xi) = -k_BT\ln
      \big<\delta\left[\xi - \sigma(\rr)\right]\big>,\;
\end{equation}
where the angular brackets denote an ensemble average. The free energy
$f(\xi)$ (sometimes also referred to as the potential of mean force)
describes the relative stability of states corresponding to
different values of $\xi$, and can provide insight into
the possible transitions between these states.

In the \abbr{ABMD} method an additional time-dependent biasing potential
is added to the potential energy of the system
\begin{equation}
  \nonumber
  \Phi_{\abbr{ABMD}}\left(\rr,t\right)
      = \Phi\left(\rr\right) +
    U\big[\sigma\left(\rr\right),t\big].
\end{equation}
This biasing potential ``floods'' the true free energy landscape as it
evolves in time according to:
\begin{equation}
  \nonumber
  \frac{\partial U(\xi,t)}{\partial t} = \frac{k_B T}{\tau_F}
    G\big[\xi - \sigma\left(\rr\right)\big].
\end{equation}
In the last equation $G(\xi)$ denotes the \emph{kernel},
which is thought to be positive and symmetric (in analogy to the
\textit{kernel density estimator} widely used in
statistics\cite{Silverman_Density_Estimation}).
For large enough $\tau_F$ (flooding timescale) and small enough
kernel width, the biasing potential $U(\xi,t)$ converges towards
$-f(\xi)$ as $t\to\infty$\cite{Lelievre_T_2007,Bussi_G_06}.

The implementation of the method is straightforward: we use cubic
B-splines\cite{book:splines} (or products of thereof for multidimensional
collective variables) to discretize $U(\xi,t)$ in ``space'':
\begin{equation}
  \nonumber
  U(\xi,t) = \sum\limits_{m} U_m(t)B(\xi/\Delta\xi - m),
\end{equation}
\begin{equation}
  \label{eq:basis}
  B(\xi) = \left\{
    \begin{array}{ll}
      (2 - |\xi|)^3/6, & 1\leq |\xi| < 2, \\
      \xi^2(|\xi| - 2)/2 + 2/3, & 0\leq |\xi| < 1, \\
      0, & \mathrm{otherwise}.
    \end{array}
  \right.
\end{equation}
We choose the biweight kernel\cite{Silverman_Density_Estimation} for
$G(\xi)$:
\begin{equation}
  \nonumber
  G(\xi) = \frac{48}{41}\left\{
    \begin{array}{ll}
      \left(1 - \left.\xi^2\right/4\right)^2, & -2 \leq \xi \leq 2 \\
      0, & \mathrm{otherwise}
    \end{array}
  \right.,
\end{equation}
and Euler-like discretization scheme for the time evolution
of the biasing potential:
\begin{equation}
  \nonumber
  U_m(t + \Delta t) = U_m(t)
  + \Delta t\frac{k_BT}{\tau_F}G\big[\sigma/\Delta\xi - m\big],
\end{equation}
where $\sigma = \sigma(\rr)$ is at time $t$. This
discretization in time may readily be improved. However, this
is not really important, since the goal is simply to 
flatten $U(\xi,t) + f(\xi)$ in the long time limit.
\subsection{Combination of replica exchange and \abbr{ABMD}}
In this paper the \abbr{ABMD} method is used to construct the
Hamiltonians for the replica exchange scheme. In particular, we are
interested in the properties of a relatively small solute, with a few
degrees of freedom, immersed in a much larger ``bath''
of solvent molecules. Parallel tempering is impractical for such a
system due to the large total number of degrees of freedom, requiring
many replicas that spend most of the time
 equilibrating the solvent at different
temperatures. On the other hand, especially for small solutes,
one can readily identify the slowest modes that are difficult to sample
in regular \abbr{MD} on the grounds of physical and/or chemical
intuition. Quite often those modes can be described as a motion along a
collective variable $\sigma = \sigma(\rr_1,\dots,\rr_S)$ that does not
depend on the solvent coordinates explicitly. It is therefore reasonable
to consider two (or more -- see below) replicas, both running at the
same temperature, with the potential energies
\begin{equation}
  \nonumber
  \Phi_a(\rr) = \Phi(\rr),\quad
  \Phi_b(\rr) = \Phi(\rr) - f\big[\sigma(\rr_1,\dots,\rr_S)\big],
\end{equation}
where the last term in $\Phi_b(\rr)$ is the Landau free energy associated
with the collective variable $\sigma$ computed for the original potential
energy $\Phi(\rr)$. This last term renders the states with
the different values of the collective variable equally probably in
the second replica, which thus plays the role of the high-temperature
replicas in the parallel tempering scheme, ``speeding'' up the motions
along the collective variable.
The first replica uses the unmodified potential energy, thus producing
unbiased samples of the configurations of the original system.
Since both replicas run at the same temperature, the exchange probability
given in Eq.(\ref{eq:exchange_probability})
does not depend explicitly on the potential energy difference between
the replicas, but on the biasing potentials as a function of the \textit{solute}
degrees of freedom. Therefore, the necessary number of replicas is expected
to be much smaller than it would be for parallel tempering.

The true free energy, $f(\xi)$, is typically unknown in advance. We
therefore use the \abbr{ABMD} method to compute a biasing potential
$U(\xi)$ that approximately compensates for the unknown free energy
$f(\xi)\approx -U(\xi)$ (up to an additive constant). This biasing potential
is then used to set up the replica exchange scheme that yields both an
enhanced sampling of the solute states and a more accurate Landau free energy
$f(\xi)$ associated with $\sigma(\rr_1,\dots,\rr_S)$. The latter is
reconstructed using the familiar \abbr{WHAM}
approach\cite{Ferrenberg_A_89,Kumar_S_92} briefly summarized in the next
subsection.

The overall success of the replica exchange scheme is determined by the
efficiency of the random walk in the replica space. The latter can be
quantified by the exchange rates between the pairs of replicas.
Let us consider two replicas running at the same temperature $T$ with
the potential energies
\begin{equation}
  \nonumber
  \Phi_{a,b}(\rr) =
    \Phi(\rr) + U_{a,b}\big[\sigma(\rr_1,\dots,\rr_S)\big],
\end{equation}
given by the original potential energy
$\Phi(\rr)$ biased by the potentials $U_{a,b}(\xi)$
acting on the collective variable $\sigma(\rr_1,\dots,\rr_S)$, which
is the same in both replicas. The exchange probability
Eq.(\ref{eq:exchange_probability}) then becomes
\begin{equation}
  \nonumber
  \mathcal{P}_{ab} = \Theta(\Delta)e^{\textstyle -\Delta}
      +\Theta(-\Delta),
\end{equation}
where $\Theta(x)$ is the Heaviside step function,
\begin{equation}
  \nonumber
  \Delta = \frac{1}{k_BT}
    \big[U_a(\xi_b) - U_a(\xi_a) + U_b(\xi_a) - U_b(\xi_b)\big],
\end{equation}
and $\xi_{a,b}$ are the values of the collective variable
$\sigma(\rr_1,\dots,\rr_S)$ in the corresponding replicas.
The probability densities of the collective variables can be
expressed through the associated Landau free energy, $f(\xi)$,
computed for the original potential $\Phi(\rr)$:
\begin{equation}
  \nonumber
  p_{a,b}(\xi)\propto
    e^{\textstyle -\big[f(\xi) + U_{a,b}(\xi)\big]\big/k_BT},
\end{equation}
and can readily be used in the expression for the rate of exchange
between the two replicas
\begin{equation}
  \nonumber
  \mathcal{R}_{ab} =
    \int\!\!\ud\xi_a\ud\xi_b\; p_a(\xi_a)p_b(\xi_b)\mathcal{P}_{ab}.
\end{equation}
The integrals in the above formula are low-dimensional and can easily be
computed numerically.

From a practical point of view, the two replicas used in the preceding
discussion often render the exchange rate unacceptably low. This can be
solved by introducing more replicas, such that the range of biasing
potentials goes from zero to the full (approximate) negative free energy
through some intermediates. For the present study we chose to use three
replicas biased as follows:
\begin{align*}
  &\Phi_1(\rr) =
    \Phi(\rr),\\
  &\Phi_2(\rr) =
    \Phi(\rr) + \lambda U\big[\sigma(\rr_1,\dots,\rr_S)\big],\\
  &\Phi_3(\rr) =
    \Phi(\rr) + \;\;U\big[\sigma(\rr_1,\dots,\rr_S)\big],
\end{align*}
where the parameter $\lambda$ is found by imposing equality of exchange rates,
$\mathcal{R}_{12} = \mathcal{R}_{23}$. This equation is solved numerically. The resulting
exchange rates turned out to be around 30\% and were deemed satisfactory.
Otherwise, more intermediate replicas would have to be introduced following
the same reasoning.
\subsection{\label{sec:wham}
Weighted histogram analysis method
}
Here we review the \abbr{WHAM}\cite{Ferrenberg_A_89,Kumar_S_92} scheme,
following the presentation given in
Refs.\onlinecite{Bartels_C_97,Bartels_2000,Habeck:200601}.
The aim of the \abbr{WHAM} approach is to infer the \textit{unbiased}
free energy out of several \textit{biased} simulations. More precisely, let
us consider $M$ simulations biased by some potentials $U_m$, $m=1,\dots,M$
that depend on atomic positions through a collective variable
$\sigma(\rr)$, that is,
\begin{displaymath}
  U_m = U_m\big[\sigma(\rr_1,\dots,\rr_N)\big].
\end{displaymath}
Let $K_m$ denote the number of values of the collective variable $\sigma$
collected in the $m^{th}$ simulation:
$\xi_{m,1},\xi_{m,2},\dots,\xi_{m,K_m}$ (with $m=1,\dots,M$).
By the end of the simulations, the whole sample $\mathcal{D}$ contains
$K_1 \times K_2\times \dots \times K_M$ values of the collective variable
that are supposed to be statistically independent.

The \textit{likelihood}\footnote{Likelihood, $p(A|B)$, is conditional
probability of $A$ given $B$.} to observe a single value
of the collective variable in the $m^{th}$ simulation is
\begin{equation}\label{eq:likelihood1}
  p(\xi|m) = \frac{1}{\mathcal{Z}_m}\int\!\!\ud\rr\;
    \delta\big[\xi - \sigma(\rr)\big]e^{\textstyle-\beta\Phi_m},
\end{equation}
where $\beta = 1/k_BT$, $\Phi_m = \Phi + U_m$ and
\begin{equation}
  \nonumber
  \mathcal{Z}_m = \int\!\!\ud\rr\; \mathrm{e}^{\textstyle-\beta\Phi_m}.
\end{equation}
The likelihood (\ref{eq:likelihood1}) can be expressed through
the Landau free energy $f(\xi)$ (which is to be determined) associated with
the collective variable
\begin{equation}
  \nonumber
  \mathrm{e}^{\textstyle -\beta f(\xi)} = \frac{1}{\mathcal{Z}}
    \int\!\!\ud\rr\;\delta\big[\xi - \sigma(\rr)\big]
    \mathrm{e}^{\textstyle-\beta \Phi},
\end{equation}
as
\begin{displaymath}
  p(\xi|m) = \mathrm{e}^{\textstyle-\eta_m -\beta f_m(\xi)},
\end{displaymath}
where
\begin{displaymath}
  \mathrm{e}^{\textstyle \eta_m}
    = \frac{\mathcal{Z}_m}{\mathcal{Z}}
    = \int\!\!\ud\xi\;\mathrm{e}^{\textstyle -\beta f_m(\xi)},
\end{displaymath}
\begin{displaymath}
  f_m(\xi) = f(\xi) + U_m(\xi),
\end{displaymath}
and
\begin{displaymath}
  \mathcal{Z} = \int\!\!\ud\rr\;\mathrm{e}^{\textstyle-\beta\Phi}.
\end{displaymath}
The likelihood of the whole dataset $\mathcal{D}$ is
\begin{equation}\label{eq:likelihoodD}
  L(f|\mathcal{D}) =
    \prod\limits_{m=1}^{M}\prod\limits_{j = 1}^{K_m} p(\xi_{j, m}|m).
\end{equation}
By introducing individual ``densities''
\begin{equation}\label{eq:distribution}
  \Gamma_m(\xi) = K_m^{-1}\sum\limits_{j=1}^{K_m}\delta(\xi - \xi_{j, m}),
\end{equation}
one can rewrite the likelihood (\ref{eq:likelihoodD}) as follows
\begin{align}\label{eq:likelihood}
  -\ln L(f|\mathcal{D}) =& \\
  \nonumber
    \sum\limits_{m=1}^{M}K_m&\int\!\!\ud\xi\;\Big[\eta_m
      + \beta f_m(\xi)\Big]\Gamma_m(\xi).
\end{align}
The maximum likelihood principle
\begin{equation}
 \label{eq:maximum-likelihood}
 \frac{\delta}{\delta f(\xi)} L(f|\mathcal{D}) = 0
\end{equation}
then leads to the celebrated \abbr{WHAM} equations\cite{Kumar_S_92}
\begin{equation}\label{eq:wham}
 \mathrm{e}^{\textstyle -\beta f(\xi)} =
   \frac{\sum\limits_{m=1}^M w_m\Gamma_m(\xi)}{\sum\limits_{m=1}^{M}
     w_m\mathrm{e}^{\textstyle -\eta_m -\beta U_m(\xi)}},
\end{equation}
where
\begin{displaymath}
   w_m = \frac{K_m}{\sum\limits_{m=1}^{M} K_m}.
\end{displaymath}
In practice\cite{Ferrenberg_A_89,Kumar_S_92,Roux_B_95}, people typically
use histograms to approximate the densities Eq.(\ref{eq:distribution}) and
then solve the \abbr{WHAM} equations Eq.(\ref{eq:wham}) to find the weights
$\eta_m$ iteratively.

Another possibility, which is exploited in this paper, is to
maximize the likelihood with respect to the unknown free energy $f(\xi)$
numerically. There are some issues with this approach too. First, the
method of maximum likelihood is not very useful if applied directly to the
likelihood given by Eq.(\ref{eq:likelihood})
because the functional is maximized by a sum of delta function spikes
at the observations\cite{Silverman_Density_Estimation}.
 The likelihood should be maximized over a particular
class of functions instead. Technically this is typically implemented
by adding a penalty term to the likelihood\cite{Good_1971,Silverman_1982b}.
While theoretically interesting, 
in this paper we overcome the problem in a similar way that \abbr{WHAM}
does: by smoothening the ``densities'' (\ref{eq:distribution}). Unlike the
classical \abbr{WHAM} papers, where the densities are approximated by
piece-wise constants (histograms), we use the so-called \textit{kernel
density estimator} for the biased densities (\ref{eq:distribution}).
That is, we replace $\Gamma_m(\xi)$ by $\widetilde{\Gamma}_m(\xi)$ which
is obtained from the former by using Gaussian kernels 
\begin{equation}\label{eq:kernel}
   \mathcal{G}(\xi) =
     \frac{1}{h\sqrt{2\pi}}\exp\left[-\frac{1}{2}\xi^2/h^2\right]
\end{equation}
in place of the $\delta$-functions.

Another issue with the functional (\ref{eq:likelihood}) is that it is
invariant with respect to the shifts of the free energy $f(\xi)$ by a constant.
While perfectly physical, this feature needs special attention in a numerical
setting and hence it is desirable to get rid of it before hand.
To this end, we consider the following
functional\cite{Silverman_Density_Estimation,Silverman_1982b}
in place of the original one:
\begin{align}\label{eq:action}
  \widetilde{L} &= \mathrm{e}^{\textstyle w_1\eta_1+\dots+w_M\eta_M}\\
  \nonumber
  +& \beta\sum\limits_{m=1}^{M}w_m\int\!\!d\xi f(\xi)\Gamma_m(\xi).
\end{align}
It is easy to see that the unconstrained minimum of the $\widetilde{L}$
functional coincides with a constrained maximum of the original
likelihood (\ref{eq:likelihood}). The functional (\ref{eq:action}) is
thus much more convenient for numerical treatment since
the constrained optimization problem gets reduced to an unconstrained
one.
\begin{figure}
\centering
\includegraphics[scale=1.0]{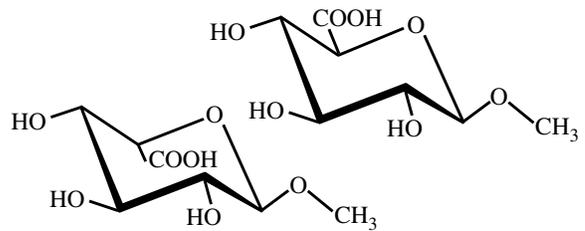}
\caption{\label{fig:epimers}
Structure of the methyl glycosides of $\alpha$-L-iduronic
acid (IdoA, left) and its C5 epimer $\beta$-D-glucuronic acid (GlcA, right).
}
\end{figure}

Technically, we represent the unknown free energy $f(\xi)$ in the
cubic $B$-splines\cite{book:splines} basis, just as we do with the
biasing potential $U(\xi)$. The functional
$\widetilde{L}$ then becomes a function of the coefficients. Finally,
we minimize this function to find the unknown coefficients
numerically using the \abbr{L-BFGS} algorithm\cite{Nocedal_J_1980,Liu_D_1989}.
%
\section{\label{sec:chemistry}
$\alpha$-L-iduronic and $\beta$-D-glucuronic methyl glycosides
}
%
\begin{figure*}
\centering
\includegraphics[scale=1.0]{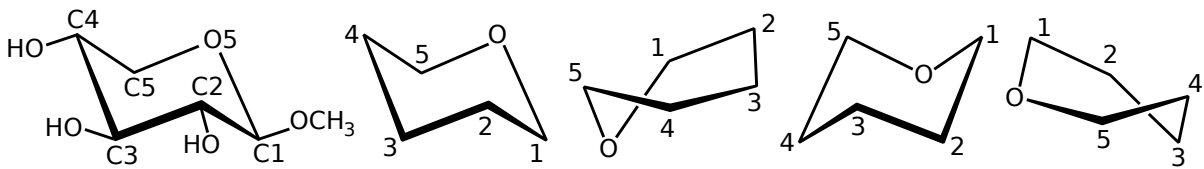}
\caption{\label{fig:conformations}
Ring carbon atom numbering (ring oxygen atom O5 is the atom
number zero) and some conformers of the pyranoid ring:
$^4C_1$, $^2S_0$, $^1C_4$ and $^1S_3$ (from left to right).
}
\end{figure*}
We consider methyl glycosides of the $\alpha$-L-iduronic acid (IdoA),
the major uronic acid component of the glycosaminoglycans dermatan
sulfate and heparin\cite{Essentials_of_Glycobiology}, and its C5 epimer,
$\beta$-D-glucuronic acid, predominant in heparan
sulfate (Fig.\ref{fig:epimers}). The IdoA residue is special among
pyranoses because it can adopt several conformations, thus giving
flexibility and conformational freedom to the corresponding
polysaccharide chains\cite{Ferro_1990}. We concentrate on
the conformational sampling of these molecules as parameterized by the
\abbr{GLYCAM~06}\cite{glycam06} force-field in explicit
\abbr{TIP3P}\cite{Jorgensen_W_83} water.

There are fourteen ``canonical'' puckering states of the pyranose
ring that minimize the angle strain: two chairs, six boats and
six skew-boats (see Ref.\onlinecite{book:Stoddart} for a more comprehensive
presentation). The $^4C_1$ and $^1C_4$ chair conformations,
the $^2S_0$ and $^1S_3$ skew-boats, and the nomenclature for
the ring carbon atoms, are
sketched in Fig.\ref{fig:conformations}. The puckering states are named
with a $C$, $B$ or $S$ letter for chair, boat or skew-boat
respectively. The letter is then superscripted/subscripted
with the number(s) of the atom(s) above/below the reference plane
(the superscripts are placed in front of the letter while the
subscripts go after it). The reference plane is chosen to contain
four of the ring atoms and if ambiguity occurs (in the chair
and skew-boat conformers), the plane is chosen so that the lowest
numbered carbon atom is out of this plane.

For the majority of pyranoses, the $^4C_1$ chair
conformation is the only relevant one with the others being much
higher in energy. The IdoA is an exception for which both
$^4C_1$ and $^1C_4$ chairs have been observed\cite{Ferro_1990}
along with the $^2S_0$ skew-boat.
Given the special place of the iduronic acid among the rest of
the pyranoses, it is important to be able to accurately reproduce
its conformational equilibrium in molecular dynamics simulations.
However, the barriers separating
the two chair conformations are of order of at least
$\approx 10 k_BT$\cite{Forster_M_93,Krautler_V_07}, and thus the
corresponding transitions are rather unlikely on the 10-100 nanosecond
timescale. In particular, in Ref.\onlinecite{Forster_M_93},
the authors conclude: ``The frequency of ring conformation
interchange ... is too slow to allow the accurate prediction of the
proportion of each form from molecular dynamics alone''.
The combination of the methods described in the preceding section
allows us to overcome this difficulty.
%
\section{\label{sec:results}
Results
}
%
\begin{figure*}
\centering
\includegraphics[width=\linewidth]{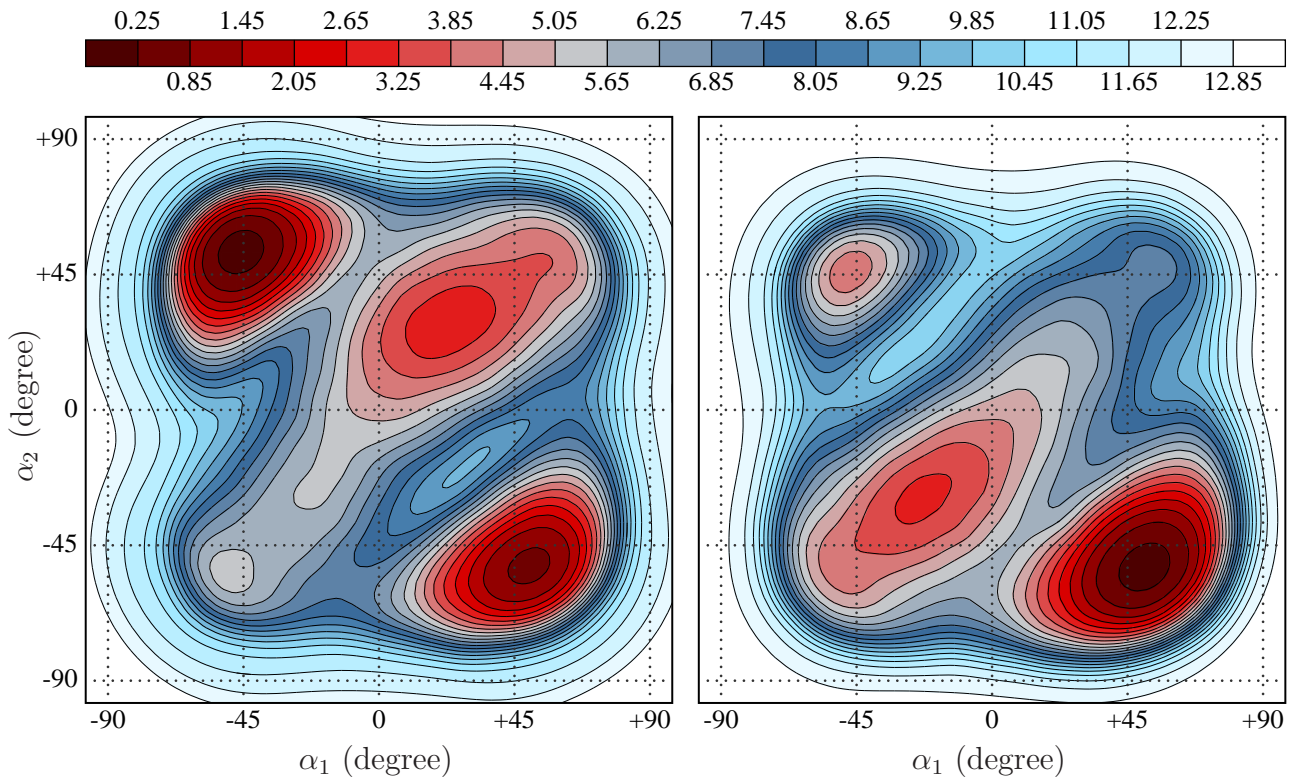}
\caption{\label{fig:fe}
Free energies (kcal/mol) associated with the O5--C1--C2--C3 ($\alpha_1$)
and C3--C4--C5--O5 ($\alpha_2$) dihedral angles for the
IdoA (left) and GlcA (right) methyl glycosides (shown
in Fig.\ref{fig:epimers}) in aqueous solution (\abbr{TIP3P} water).
}
\end{figure*}
We use the intra-ring O5--C1--C2--C3 and C3--C4--C5--O5 torsion angles
(denoted $\alpha_1$ and $\alpha_2$ in the following) to distinguish
among the different conformations. Our choice is not unique, but this is
of minor importance since the dihedrals adopt distinctly different values
for the conformations in question: the values of the angles are of
different sign for the chairs (with $^1C_4$ corresponding to
$\alpha_1\approx -50^\circ$, $\alpha_2\approx +50^\circ$ and $^4C_1$
having $\alpha_1\approx +50^\circ$, $\alpha_2\approx -50^\circ$) and  of
the same sign for the boats and skew-boat conformations.

Our goal is to sample the equilibrium configurations of the rings in the
\abbr{NTV} ensemble using explicit solvent. We proceed in
stages as follows: (1) compute
with \abbr{ABMD} an approximate free energy for the dihedral angles
using a relatively cheap and fast implicit solvent model; (2) introduce explicit
solvent and ``refine'' the potential of mean force (\abbr{PMF}) from the
previous step; (3) set up a replica exchange scheme that uses the
approximate \abbr{PMF} to sample from an unbiased canonical distribution while
collecting the biased samples as well. The trajectories from the
last step are then used to study equilibrium properties of the rings and
to recover the accurate free energies using the \abbr{WHAM} technique.
The general philosophy behind the protocol is to start with coarser,
considerably cheaper approximations that are successively refined with
more accurate, expensive approaches such that the final results have the
accuracy of the expensive approach at a fraction of the cost. Thus
we start with implicit solvent and then use these results as the
starting point for explicit solvent.

Let us describe the simulation protocol step by step. The simulation
technical details can be found in Appendix~\ref{sec:numerics}.
\subsection{Step~1: flooding with implicit water}
We start sampling in implicit water. A short run at $T=300K$ revealed
that the angles fluctuate with an amplitude of $\approx 10$ degrees and
a characteristic time (taken to be the period of oscillations here) of
0.5ps. We thus performed a flooding \abbr{ABMD} run with $4\Delta\xi =
17^\circ$ (for both torsional angles) and $\tau_F = 1\mathrm{ps}$. We
monitored the trajectory and continued running until both angles visited
all the values between $\approx -80^\circ$ and $\approx +80^\circ$ (that
is, until the minima corresponding to both the $^4C_1$ and $^1C_1$ chair
conformations and skew-boat configurations were ``flooded''). The coarse
stage took 1ns. We then decreased the spatial resolution to $4\Delta\xi
= 11^\circ$ and increased the timescale to $\tau_F = 50\mathrm{ps}$ and
continued the flooding to refine the biasing potential obtained in
the previous step. It took 3ns more of simulation time
 to cover the desired values of
$\alpha_{1,2}$. The biasing potential at this stage is expected to
provide a relatively good approximation of the corresponding free
energy. (It is not necessary to compute the error at this stage since
this potential is only an approximation to be used in the next step).

Regarding the choice of the ``spatial'' resolution $4\Delta\xi$, we
have to balance two conflicting factors: a smaller resolution allows the
identification of the finer details of the free energy; yet for a
smaller $4\Delta\xi$ (and fixed flooding rate $\tau_F$), the accuracy of
the free energy decreases because the system experiences stronger biasing
forces pushing it out of the thermodynamic equilibrium. We thus choose
the resolution to be sufficiently small to approximate the equilibrium
minima well.
\subsection{Step~2: flooding with explicit water}
It is reasonable to expect that the biasing potential obtained with
implicit solvent provides a good approximation for the free energy in
explicit solvent. To account for the differences between the two solvent
descriptions, we carried out more flooding in explicit solvent with
$\tau_F = 50\mathrm{ps}$ and $4\Delta\xi = 11^\circ$ (i.e., the ``fine''
values used for the implicit solvent simulation). After just 5ns, every value
of the angles was visited and so we stopped the flooding. Next, we
performed a 5ns equilibrium \abbr{MD} run biased by the biasing
potential computed so far. The biased trajectory visited all the values
in the interval $-80^\circ\leq\alpha_{1,2}\leq +80^\circ$. This means
that the biasing potential approximates the true free energy surface to
within a few $k_BT$ (otherwise the trajectory would get trapped
somewhere; the fact that $\alpha_{1,2}$ sweep the whole region implies
that the error is of order of a few $k_BT$ at most).
\subsection{Step~3: replica exchange with explicit water}
\begin{figure*}
\centering
\includegraphics[scale=1.0]{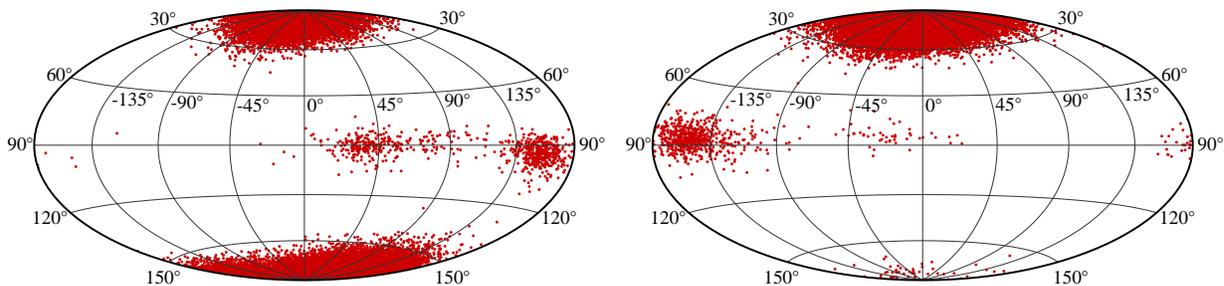}
\caption{\label{fig:pucker}
The $(\vartheta, \varphi)$ angles of pseudo-rotation\cite{Cremer_D_75} for
the IdoA (left) and GlcA (right) methyl glycosides in
aqueous solution (\abbr{TIP3P} water) at $T=300$K (using the
Hammer-Aitoff\cite{Snyder_J_1997} equal-area projection).
}
\end{figure*}
Finally, we set up a replica exchange scheme that samples both biased and
unbiased configurations. This allows for: (a) the computation of the
accurate free energy associated with the ring puckering states, and (b)
the computation of the unbiased equilibrium averages, such as
$J$-couplings, which are of particular interest since they allow for a
direct comparison with the \abbr{NMR} data. This last aspect is of
paramount importance for the force-field development.

As we described in the ``Methods'' section, we put three replicas
together: the first replica is not biased, the second one is
biased by $\lambda U(\xi)$, and the third one is
biased by the full $U(\xi)$. We determined the value of $\lambda=0.7$
numerically solving the equation $\mathcal{R}_{12}=\mathcal{R}_{23}$.
After a short 1ns equilibration, we went into ``production runs" for
100ns trying to exchange a random pair of replicas every 1ps. The
observed exchange rates of $\approx 30\%$ were in perfect agreement with
the expected values (computed assuming that $f(\xi) = -U(\xi)$) for both
IdoA and GlcA cases.

We emphasize that the simulations in explicit solvent involve jumps over
 barriers of order of 6.5kcal/mol ($\approx 10 k_BT$). In order for
conventional parallel tempering to be able to sample this system, the
highest temperature replica would have to run at $\approx 10\times
300K$. This is a prohibitively large number since several hundreds
replicas running at intermediate temperatures would be needed to proxy
between the highest and the lowest (room) temperatures.
\subsection{Accurate free energies using \abbr{WHAM}}
Since the autocorrelation time for the angles in a fully biased
\abbr{MD} run is approximately 10ps, we regarded the samples taken 10ps
apart during the equilibrium replica exchange as statistically
``independent''. Therefore we saved the coordinates from the unbiased
replica every 10ps to compute various equilibrium properties of
the IdoA and GlcA, reported in the subsection below. We used the whole
sample from all three replicas and the \abbr{WHAM} scheme
to recover the accurate free energy, shown in Fig.\ref{fig:fe}.

A crucial part of this step is the selection of the kernel bandwidth
$h$ in Eq.(\ref{eq:kernel}).
Assuming that the data is statistically independent, we
computed optimal $h$ by minimizing the least-squares cross-validation
score to get $h\approx 1.1^\circ$ for all the runs; we also computed the
bandwidths by maximizing the likelihood cross-validation score getting
$h\approx 2.9^\circ$ for all the runs. The values selected by both
methods were checked using the Silverman's ``visual'' method and
appeared to be too small (a problem described, for example,
in Ref.\onlinecite{Loader_C_1999}). Combining all the hints we chose
$h = 5.73^\circ,\;8.59^\circ\;\mathrm{and}\;11.46^\circ$ for the
unbiased, intermediate and fully biased replicas, respectively.
\subsection{Equilibrium properties of the compounds}
%
%
%
We computed the Cremer-Pople puckering parameters\cite{Cremer_D_75}
regarding the O5 oxygen as the ring apex (atom number one in the 
Cremer-Pople formulae) and the carbon with the glycosidic linkage
(C1) as the atom number two.
The mean value of the puckering amplitude $Q$ turned out to be
0.53{\AA} and 0.55{\AA} (for IdoA and GlcA respectively) with standard
deviation $\approx 0.04${\AA} for both compounds.
Both $\vartheta$ and $\varphi$ pseudo-rotation angles display somewhat
larger variations and hence are shown in Fig.\ref{fig:pucker} using the
Hammer-Aitoff\cite{Snyder_J_1997} equal-area projection:
\begin{displaymath}
  (\vartheta, \varphi)\to\left(
    \frac{2\sqrt{2}\sin\vartheta\sin\varphi/2}
         {\sqrt{1 + \sin\vartheta\cos\varphi/2}},
    \frac{\sqrt{2}\cos\vartheta}{\sqrt{1 + \sin\vartheta\cos\varphi/2}}
   \right).
\end{displaymath}
The $(\vartheta,\varphi)$ values are consistent with the free energies
associated with the $\alpha_{1,2}$ (see Fig.\ref{fig:fe}): there are two
well-pronounced clusters near the poles for IdoA corresponding to the
$^4C_1$ (north pole, $\vartheta = 0^\circ$) and $^1C_4$ (south pole,
$\vartheta = 180^\circ$) chair conformations, and just one big cluster
near the north pole ($^4C_1$ chair) for the GlcA. The boat and twisted
boat conformations correspond to the equatorial area ($\vartheta = 0^\circ$)
and are indeed considerably less frequent than the chairs. For IdoA
the majority of the ``equatorial''
samples is around $\varphi = 150^\circ$ ($^2S_0$ skew-boat),
yet several other twisted boat conformations are also present. This
is in accord with Ref.\onlinecite{Ernst_S_98} where it has been suggested
that not only ``canonical'' boat and skew-boat conformations, but also
various boat-like intermediates, contribute to the conformational
equilibrium. The most frequent skew-boat conformation for GlcA observed
to be $^1S_3$ ($\varphi$ around $-150^\circ$).

In order to identify the pyranose puckering states 
 explored during the simulations,
one has to choose a criterion for assigning a particular trajectory section
to a canonical puckering state. When the temperature is zero, the
potential energy minima typically define the conformations in a unique and natural
way. At non-zero temperature, however, some ambiguity is always present
due to the thermal fluctuations.
In our simulations, the $^4C_1$ and $^1C_4$ chair conformations are clearly
discriminated by the signs of the $\alpha_{1,2}$ angles and hence we
assume that the equilibrium samples with $(\alpha_1>0,\alpha_2<0)$ are
$^4C_1$ chairs while those with $(\alpha_1<0,\alpha_2>0)$ are $^1C_4$
chairs. The boat and skew-boat conformations, on the other hand, correspond to
angles of the same sign, yet it is difficult to distinguish among
the different boat and skew-boat types using the $\alpha_{1,2}$ angles alone.
The Cremer-Pople parameters could potentially discriminate among the boats,
but the ambiguity still persists as one would have to set the
boundaries between the conformations. Thus, in the following we concentrate 
on the properties of the chair conformations, although our simulations
show that different skew-boat conformations show up at the thermodynamic
equilibrium. In the case of IdoA the majority of those are $^2S_0$-like
(left panel in Fig.\ref{fig:pucker}: dots in the equatorial area
with $\varphi$ around $150^\circ$) and $^3S_1$-like
($\varphi$ around $30^\circ$). This
is in agreement with the conclusions of Mulloy and Forster\cite{Mulloy_B_00} 
that the $^4C_1\leftrightarrow ^1C_4$ transformation of IdoA proceeds through
the $^2S_0$ and $^3S_1$ intermediates.

We were able to compute the free energy differences between the
$^4C_1$ and $^1C_4$ states by counting the number of configurations 
$\mathcal{N}(\alpha_1,\alpha_2)$ with
the appropriate signs of the $\alpha_{1,2}$ angles:
\begin{align*}
\Delta G^{\circ} &= -k_BT\ln\frac{p(^4C_1)}{p(^1C_4)}\\
  &\approx -k_BT\ln\frac{\mathcal{N}(\alpha_1>0,\alpha_2<0)}{\mathcal{N}(\alpha_1<0,\alpha_2>0)},
\end{align*}
where $p(^4C_1)$ and $p(^1C_4)$ denote the probabilities of the
corresponding conformations. We obtained
\begin{displaymath}
  \Delta G^{\circ}_{\mathrm{IdoA}} = 0.71\mathrm{kcal/mol},\;
  \Delta G^{\circ}_{\mathrm{GlcA}} = -4.49\mathrm{kcal/mol}.
\end{displaymath}
In other words,  $^1C_4$ turns out to be marginally more stable
than $^4C_1$ for IdoA, while for GlcA,  $^4C_1$ is considerably
more stable than $^1C_4$. The preference of $^1C_4$ over $^4C_1$
for IdoA appears to be largely due to solvation effects -- the vacuum
energies optimized in the \abbr{GLYCAM~06} force-field differ merely by
0.2kcal/mol. On the other hand, the $^4C_1$ has much lower energy
than $^1C_4$ for the GlcA in the gas phase and thus the $^4C_1$
stability in water is more due to the ``internal'' energetics than
to solvent interactions or temperature.
\begin{table}
\begin{tabular}{l|cc|c}
\hline
  & IdoA & & GlcA \\
\cline{2-4}
  & $^4C_1$ & $^1C_4$ & $^4C_1$ \\
\hline
Bond angle & & & \\
C5--O5--C1 & 116.7(3.3) & 116.6(3.3) & 116.0(3.5) \\
O5--C1--C2 & 110.0(3.0) & 111.4(2.8) & 109.7(3.0) \\
C1--C2--C3 & 111.6(3.2) & 112.8(2.8) & 112.1(3.2) \\
C2--C3--C4 & 113.4(3.2) & 114.2(3.0) & 112.6(3.2) \\
C3--C4--C5 & 112.4(3.0) & 111.6(2.9) & 112.8(3.2) \\
C4--C5--O5 & 107.9(2.8) & 109.5(2.8) & 108.3(2.9) \\

 & & & \\
Dihedral angle & & & \\
C5--O5--C1--C2 & -57.9(6.7) &  53.4(6.4) & -58.2(7.4) \\
O5--C1--C2--C3 &  49.3(7.3) & -45.6(6.2) &  50.2(7.4) \\
C1--C2--C3--C4 & -47.2(7.3) &  45.3(6.3) & -47.1(7.3) \\
C2--C3--C4--C5 &  49.4(6.6) & -49.0(5.9) &  48.8(6.9) \\
C3--C4--C5--O5 & -51.8(6.0) &  52.2(5.6) & -52.0(6.7) \\
C4--C5--O5--C1 &  58.7(6.1) & -56.9(6.3) &  58.9(7.3) \\
\end{tabular}
\caption{\label{tbl:torsions}
Mean value and standard deviation (in parenthesis) of the intra-ring
bond and dihedral angles for IdoA and GlcA methyl glycosides in
$^4C_1$ and $^1C_4$ conformations (degrees).
}
\end{table}

\begin{figure}
\centering
\includegraphics[scale=1.0]{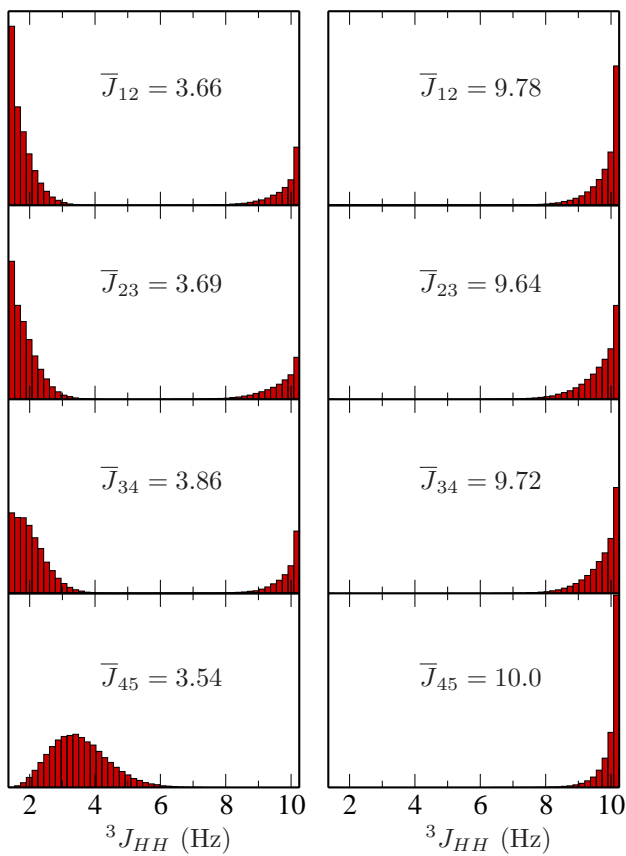}
\caption{\label{fig:jcouplings}
Distributions of the H-H coupling constants computed using
the Karplus equation for IdoA (left) and GlcA (right). Mean
value shown in the insets.
}
\end{figure}

Finally, in order to make connection with \abbr{NMR} spectral
data\cite{Ferro_1990}, we computed the J-couplings for these molecules.
Also known as indirect dipole-dipole coupling, a J-coupling is the
coupling between two nuclear spins caused by bonding electrons acting on
the magnetic field through the two nuclei. J-couplings can be linked to
the dihedral angles via  the Karplus equation\cite{Karplus_M_63}:
\begin{equation}\label{eq:karplus}
  ^3J_{HH} = A\cos^2\varphi + B\cos\varphi + C,
\end{equation}
We used the values of the parameters reported in Ref.\onlinecite{Haasnoot_80}.
The $^3J_{HH}$-coupling constants are presented in Fig.\ref{fig:jcouplings}.
We have carried out a
considerably accurate sampling of phase space for these molecules, but
naturally the validity of the final numbers depend on the validity of
the force-field and is also limited by the accuracy of the Karplus
equation Eq.(\ref{eq:karplus}). The distribution of the $^3J_{HH}$ values
for IdoA clearly reflects the contributions from the two chair
conformations, yet the mean values do not match the ones reported in
Ref.\onlinecite{Ferro_1990} exactly. Specifically, the mean values of
the $^3J_{HH}$ for the methyl-$\alpha$-L-Iduronic acid reported in
Ref.\onlinecite{Ferro_1990} (line 1, Table I in that reference) are
$J_{1,2}=4.9$, $J_{2,3}=6.6$, $J_{3,4}=6.0$ and $J_{4,5}=4.0$ Hz
implying that $^4C_1$ is favored over $^1C_4$ contrary to our results.
We also note that, due to the bimodal nature
of the IdoA's $^3J_{HH}$ distributions (see Fig.\ref{fig:jcouplings}),
the discrepancies of the mean values can be deceptive.
Unfortunately,
while there are ample experimental data on oligo- and polysaccharides,
information about their monosaccharide units is sketchy and scattered
across the scientific literature\cite{Krautler_V_07}, which prevents
more rigorous comparisons.
%
\section{\label{sec:summary}
Summary
}
%
In this work we carried out simulations of
methyl-$\alpha$-L-iduronic and methyl-$\beta$-D-glucuronic acids
parametrized by the \abbr{GLYCAM~06} force field
in explicit \abbr{TIP3P} water. The importance of these small sugars
is greatly due to the fact that IdoA is the major uronic acid
component of the glycosaminoglycans dermatan sulfate and
heparin, while GlcA  dominates in heparan sulfate. The main purpose
of this work was to introduce a procedure that reliably recovers the
free energy of the solvated molecules and samples the solute degrees
of freedom in explicit solvent using very moderate computational resources. 

Succinctly, the protocol involves the following steps: (i) use
\abbr{ABMD} to compute an approximate free energy for the intra-ring
dihedral angles in a relatively cheap implicit solvent; (ii) refine the
free energy with \abbr{ABMD} under explicit solvent; 
(iii) use this last free energy to
set up a Hamiltonian replica exchange scheme that samples both from
biased and unbiased distributions; and (iv) recover the accurate free
energies via the \abbr{WHAM} technique applied to all the replicas, and
compute equilibrium properties of the rings from the unbiased trajectories.

The protocol appears to be slightly more complex than the standard parallel
tempering method. However, big savings in computational resources fully compensate
for this extra ``complication". The simulations in explicit solvent reported
in this paper involve jumps over barriers of order of $\approx 10 k_BT$.
An efficient parallel tempering setup in this case would have to
run the highest temperature replica at $\approx 10\times 300K$. This
rather high temperature would in turn require several hundred
replicas to mediate between this highest and the room temperature.
In our scheme all
replicas run at the same temperature, and hence the exchange probability
given in Eq.(\ref{eq:exchange_probability})
does not depend explicitly on the potential energy difference between
the replicas; it depends, instead, on the differences of the biasing
potentials, which are functions of the \textit{solute} degrees of freedom
only. Therefore, the necessary number of replicas is considerably smaller
(three replicas were enough for the pyranose rings considered in this paper).

Once that accurate sampling is not an issue anymore, one can really concentrate
on problems associated with the validity of the force-field representation.
In particular, one can compare against experimentally measurable quantities,
such as $J$-couplings, with the certitude that any deviations from experimental
values are due to inaccuracies of the force-field (of course, assuming that the
experimental values are correct). Naturally, one can
extend this study to take into account the effects of different densities,
salt concentrations, temperature, etc.
%
\acknowledgments
This research was supported by NSF under grant 
FRG-0804549. We also thank the NC State HPC for computer time.
%
%
\appendix
%
%
\section{\label{sec:numerics}
Simulation details.
}
%
%
The simulations presented in this paper were performed using the
\abbr{AMBER~10}\cite{Amber10} simulation package. The \abbr{LEaP} program
from that package was used to prepare the initial structures along with
the topology/parameter files. The \abbr{GLYCAM~06}\cite{glycam06}
force field was used for the solutes.

The implicit solvent simulations were carried out using the generalized
Born model\cite{Onufriev00,Onufriev04},
including surface area
contribution computed using the \abbr{LCPO} model\cite{Weiser99} with the
surface tension set to $5\times 10^{-3}$ kcal/mol/{\AA$^2$}.
The \abbr{GB/SA} simulations were performed enforcing no cutoff on
the non-bonded (van der Waals and electrostatics) interactions.

The \abbr{TIP3P} water model\cite{Jorgensen_W_83} was used for the
explicit solvent simulations (in each case 645 water molecules were added
along with a $\mathrm{Na}^{+}$ ion to neutralize the system).
For the explicit water simulations, we used periodic boundary conditions
with fixed box size (27{\AA} in all three directions) chosen to render
the density equal to $\approx 1\mathrm{g}/\mathrm{cm}^3$. The
Particle Mesh Ewald (\abbr{PME})\cite{Darden_T_93,Essmann_U_95} method was
used with the direct space cutoff set to 9{\AA} and a $32\times 32\times 32$
grid for the smooth part of the Ewald sum.

The lengths of all bonds that contain hydrogen were fixed via the
\abbr{SHAKE} algorithm with the tolerance set to $10^{-6}${\AA}.
Langevin dynamics\cite{Brunger_A_84} with collision frequency
$\gamma=1\mathrm{ps}^{-1}$ was used to keep the temperature at 300K.
A different random number generator seed was set for every run.
%
%

%
\end{document}